\documentclass[conference]{IEEEtran}
\usepackage[utf8x]{inputenc}

\usepackage{cite}

\ifCLASSINFOpdf
  \usepackage[pdftex]{graphicx}
  \graphicspath{{./figures/}}
\else
  \usepackage[dvips]{graphicx}
  \graphicspath{{./eps/}}
\fi
\usepackage{amsmath}
\usepackage{enumerate}

\usepackage{array}

\usepackage{tabularx}
\usepackage{tabto}

\ifCLASSOPTIONcompsoc
 \usepackage[caption=false,font=normalsize,labelfont=sf,textfont=sf]{subfig}
\else
  \usepackage[caption=false,font=footnotesize]{subfig}
\fi

\usepackage[]{footmisc}
\usepackage{url}

\begin{document}
%
\title{Using Ultra-Wideband Technology in Vehicles for Infrastructure-free Localization}

\author{\IEEEauthorblockN{Rusheng Zhang \IEEEauthorrefmark{1}, Lin Song \IEEEauthorrefmark{1}, Adhishree Jaiprakash \IEEEauthorrefmark{1}, Timothy Talty  \IEEEauthorrefmark{2}, \\ Ammar Alanazi \IEEEauthorrefmark{3}, Abdullah Alghafis \IEEEauthorrefmark{3}, A. Ahmed Biyabani \IEEEauthorrefmark{4},  and Ozan Tonguz \IEEEauthorrefmark{1}}\\
\IEEEauthorblockA{\IEEEauthorrefmark{1} Department of Electrical and Computer Engineering,
Carnegie Mellon University,
 Pittsburgh, PA 15213-3890, USA}
 \IEEEauthorblockA{\IEEEauthorrefmark{2} General Motors Research \& Development, Warren, Michigan 48092, USA}
  \IEEEauthorblockA{\IEEEauthorrefmark{3} King Abdulaziz City for Science and Technology (KACST),  Riyadh, Saudi Arabia}
  
  \IEEEauthorblockA{\IEEEauthorrefmark{4} Hawaz Inc.,  Riyadh, Saudi Arabia}

}


%


\IEEEoverridecommandlockouts
\IEEEpubid{\makebox[\columnwidth]{978-1-5386-4980-0/19/\$31.00 \copyright2019 IEEE} \hspace{\columnsep}\makebox[\columnwidth]{ }}
\maketitle

\vspace{-0.5in}

\begin{abstract}




In this paper, we investigate using Ultra-Wideband (UWB) technology in vehicles for  localization as well as other possible infrastructure-free applications.

To that end, we first introduce the on-vehicle UWB anchor system configuration, then conduct a theoretical analysis to shed light on the capabilities and limitations of using UWB anchors with this configuration. Extensive field trials performed verified the validity of the analysis conducted.

Finally, Virtual Pedestrian Traffic Light (VPTL), an infrastructure-free pedestrian traffic light system is introduced as an example application of the presented approach.

\end{abstract}
\vspace{0.1in}
\emph{keywords: Ultra-Wideband, pedestrian localization, pedestrian detection, pedestrian tracking}

%
\IEEEpeerreviewmaketitle

\section{Introduction}


 Ultra-Wideband (UWB) is an emerging technology for positioning, which provides centimeter-level positioning accuracy \cite{yang2004ultra, gezici2005localization}. Recently, UWB has received a lot of attention, since it is the most promising method for indoor positioning \cite{ingram2004ultrawideband, zhou2011indoor, kok2015indoor, alarifi2016ultra}. Therefore, it is conceivable that UWB devices will be installed in cellphones and possibly other wearable devices in the future.\footnote[1]{The research reported in this paper was funded by King Abdulaziz City of Science and Technology (KACST), Riyadh, Kingdom of Saudi Arabia.}

UWB technology has  been investigated by several research groups for applications in various vehicle-related scenarios, such as parking guide \cite{tiemann2016ultra}, vehicle localization module as compensation of GPS \cite{gonzalez2007combination, fernandez2007application}, intra-vehicular networks \cite{niu2008intra}, and cyclists tracking \cite{dardari2017high}.

In this paper, we report the localization performance of using UWB  devices in vehicles, an on-vehicle anchor configuration. While this configuration has been proposed before \cite{dardari2017high}, it has not been carefully assessed, due to the fact that the localization error of this configuration is larger than other UWB anchor layouts. However, since the anchors are all installed in the vehicles, this configuration is infrastructure-free, which is clearly a significant benefit.

The main contributions of this paper are as follows:
\begin{enumerate}
    \item Providing a theoretical analysis, together with an intuitive interpretation of the error characteristics of this configuration.
    \item Reporting field test results on the error characteristics which verify the predictions of the theoretical analysis conducted.
    \item Proposing a system utilizing the in-vehicle anchor configuration, Virtual Pedestrian Traffic Light (VPTL), as an illustrative application of the developed approach.
\end{enumerate}


\vspace{-0.1in}
\section{Layout Design}

\subsection{Introduction to UWB Device}
\label{ss:intro_dev}
\begin{figure}[ht]
    \centering
    \includegraphics[width=.6\linewidth]{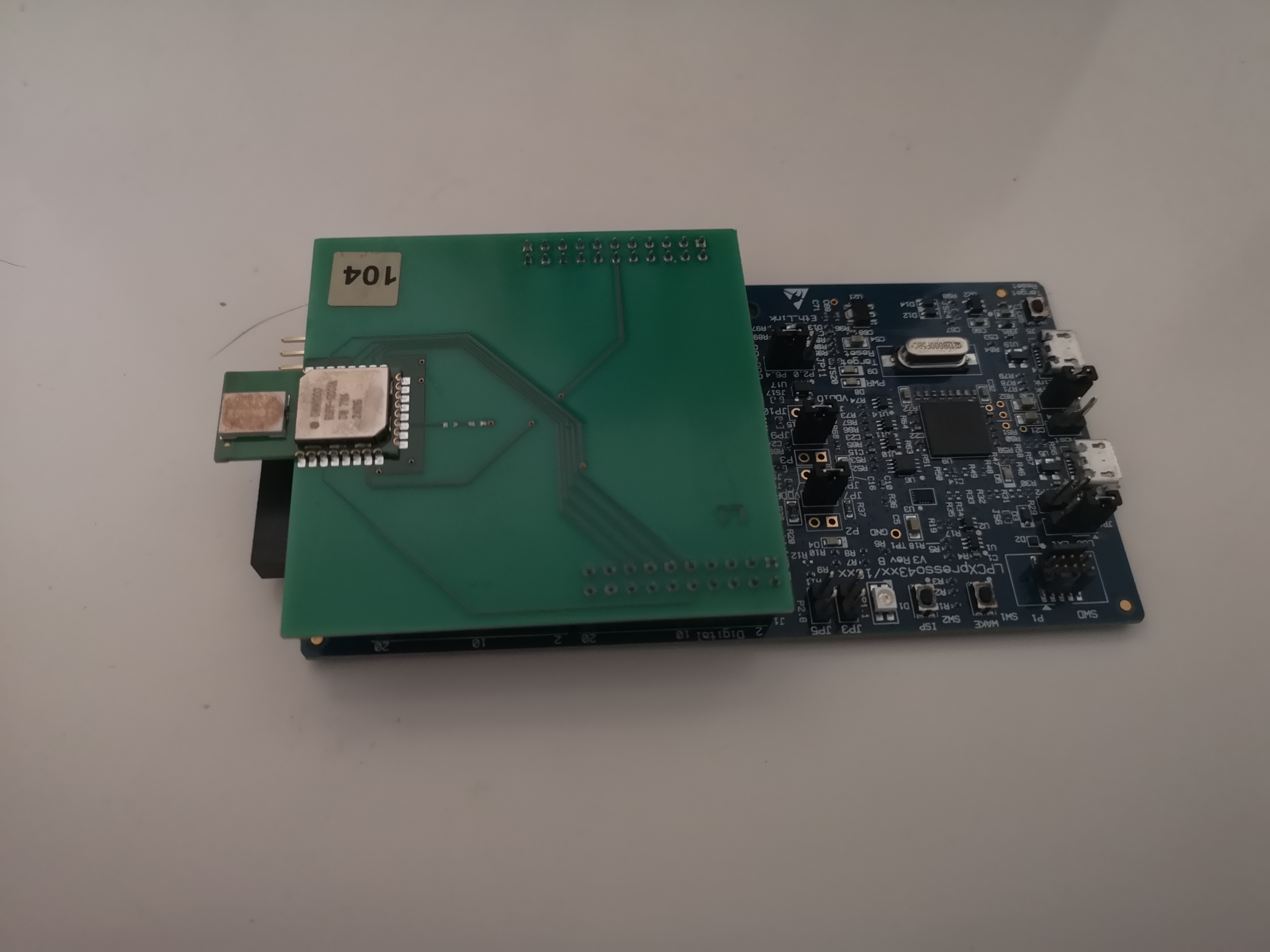}
    \caption{UWB experimental board}
    \label{fig:board}
\end{figure}

The UWB module used for the experiments conducted is a DecaWave DW1000, which is mounted on a connection board that connects to a LPCXpresso43S67 Development Board. The module is IEEE802.15.4 compliant and supports 6 RF bands from 3.5 GHz to 6.5 GHz \cite{IEEEstandard} meant for precision localization using Time Difference of Arrival (TDOA) scheme and concurrent data transfer with high multi-path fading immunity. The module is calibrated for distance measurements at room temperature for Channels 5 and Channel 2.  Figure \ref{fig:board} shows a photo of the experimental board that was used in our experiments. \footnote[2]{The experimental devices in this paper are provided by NXP Semiconductors N.V.}

The device for the experiments can be configured as either a tag or an anchor. The distance measurement is based on the time of flight between a packet sent by the tag to the anchor and a return packet received by the tag. The device offers three ranging modes - Single-Sided (SS) Two-Way Ranging, Double-Sided (DS) Two-Way Ranging and Auto -Acknowledgement (AA) Two-Way Ranging. 


SS Two-Way Ranging is initiated by the tag or anchor and expects a reply from the anchor within the set timeout. The response message is used to calculate the single-sided TDOA and thus the distance. In comparison with SS; the DS Two-Way Ranging is also initiated by the tag but ensures two-way measurement by polling the distance from the anchor as well; AA on the other hand, sends a polling message to the anchor which is acknowledged immediately by the anchor. In the field experiment, the three methods result in very similar behaviors, so in the latter part of the paper, we use DS as the distance measurement method. Since, the anchor and tag functionalities of the UWB board are interchangeable, both the anchor or the tag can be used to determine the distance between the anchor and the tag, depending on the needs of the application. In the application explored in this paper, when the vehicle tries to track the tag nearby, the anchor is the device that initiates the measurement and determines the distance.

The locations (coordinates) of the anchors are considered to be known. Therefore, by measuring the distance between the anchors and the tag, one can use triangulation techniques to calculate the coordinates of the tag. 

\subsection{On-Vehicle Anchor Layout}
Traditionally, the UWB anchors are designed to be installed on the ground, where the distance between the anchors is relatively large. Even though this layout can locate tags accurately, extra infrastructure is required to be installed at the location. However, in many vehicle-related applications, it is desirable to detect the tags in an infrastructure-free manner; for example, an autonomous vehicle has to detect pedestrians moving in front of the vehicle.  Therefore, an on-vehicle anchor layout is of great interest.

\begin{figure}[ht]
    \centering
    \includegraphics[width=.6\linewidth]{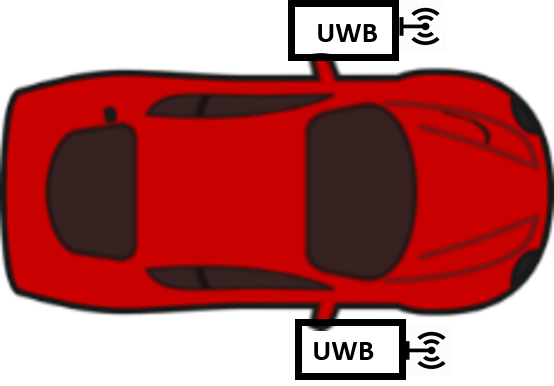}
    \caption{Conceptual figure of on-vehicle anchor configuration}
    \label{fig:layout}
\end{figure}

 Figure \ref{fig:layout} shows the on-vehicle configuration investigated in this paper.  The two UWB anchors are installed on the rear-view mirrors of the vehicle. We choose the rear-view mirrors because this choice provides the longest distance between the two anchors on the vehicle,  and hence minimizes the localization error.
 
 Many applications can benefit from this configuration, such as an in-vehicle pedestrian warning system, an infrastructure-free pedestrian traffic light (introduced in the latter part of this paper), or Passive Keyless Entry (PKE), which automatically locks and unlocks the vehicle based on the location of the key (tag) \cite{waraksa1990passive}.
 
 It is worthwhile mentioning here that to locate a tag on a plane, at least three UWB anchors are needed. Therefore, another anchor should be installed on the vehicle to select one of the two intersections of the triangulation to be the estimated location of the tag. In this paper, we evaluate the accuracy of locating using only two anchors and consider the third one only serving  an assisting role. Furthermore, since in most  applications, one is interested in positioning the tag in front of a vehicle, the error characteristics in front of the vehicle are investigated by both the theoretical analysis and the field trials conducted. 
 
 In the implementation, an computation unit on vehicle is needed, which communicate with two anchors and obtain the distances of the anchors to the tag, and compute the coordinates of the tag using triangulation.


\section{Analysis and Experiments}
\subsection{Error analysis}
\label{ss:error_analysis}
A general analysis of ranging errors of UWB stemming from Non Line-of-Sight (NLoS) conditions is reported in \cite{jourdan2008position}. In this paper, our focus is on how this error will impact the positioning accuracy of the on-vehicle layout. So, we introduce a ranging error term $e$ and see how the localization of the on-vehicle anchor configuration performs under this ranging error. In real world, this ranging error may be introduced by the NLOS or other random factors.

We use two UWB anchors and one UWB tag  to conduct the field tests. For the two anchors, suppose that we place them at the coordinates  $(-x,0)$ and $(x, 0)$ and suppose that the distance of the tag to the anchors are $r_1$ and $r_2$, respectively. The coordinates of the tag $(x_k, y_k)$ can be calculated by the following expression:

    \begin{equation}
    \begin{gathered}
        x_k = \frac{r_1^2-r_2^2}{4x}\\
        y_k = x\sqrt{\frac{r_1^2+r_2^2}{2x^2} - \frac{(r_1^2-r_2^2)^2}{16x^4}-1}
    \end{gathered}
    \label{eq:t1}
    \end{equation}

We then introduce a ranging error $e$ into the measurement of $r_1$ and $r_2$, and see how much the error $e$ will determine the change of $x_k$ and $y_k$. To get the maximum error on $x_k$,  measurement of the two distances should be $r_1+e$ and $r_2-e$, for $y_k$, to get the maximum error, the two distances should be $r_1+e$ and $r_2+e$:
\begin{equation}
\begin{gathered}
\tilde{x_k}=\frac{(r_1+e)^2-(r_2-e)^2}{4x}\\
\tilde{y_k} = x\times\Big\{
    \frac{(r_1+e)^2+(r_2+e)^2}{2x^2} - 1\\
    -\frac{[(r_1+e)^2-(r_2+e)^2]^2}{16x^4}
\Big\}^{0.5}
\end{gathered}
\label{eq:t2}
\end{equation}

The partial derivative with respect to $e$ gives the ratio of localization error to the ranging error $e$:

        \begin{equation}
        \begin{gathered}
        \frac{\partial \tilde{x_k}}{\partial e} = \frac{r_1+r_2}{2x}\\
        \frac{\partial \tilde{y_k}}{\partial e} = \frac{\frac{4e+2r_1+2r_2}{2x}-\frac{(2r_1-2r_2)\zeta}{8x^3}}{2\sqrt{\frac{(e+r_1)^2+(e+r_2)^2}{2x^2}-\frac{\zeta}{16x^4}-1}}
        \end{gathered}
        \end{equation}

where $\zeta = (e+r_1)^2-(e+r_2)^2$.

For $\tilde{y_k}$,  the partial derivative is complicated; to better understand how the error behaves,  an assumption  $r_1=r_2=r$ is made; the approximation is reasonable because under the condition $x<<r_1$ and $x<<r_2$, $r_1$ and $r_2$ are going to be very similar, and the situation $x<<r_1,r_2$ is likely to be the case for the layout to detect the tag in front of the vehicle. By assuming $r_1=r_2=r$, we will get error characteristics right in front of the vehicle. Under this assumption, the calculation of the partial derivative can be significantly simplified:

\begin{equation}
    \begin{aligned}
        \frac{\partial \tilde{x_k}}{\partial e} = \frac{r}{x}\\
        \frac{\partial \tilde{y_k}}{\partial e} = \frac{1}{\sqrt{1-\frac{x^2}{r^2}}}
    \end{aligned}
    \label{eq:1}
\end{equation}
Equation (\ref{eq:1}) gives valuable insight into the error characteristics under the proposed anchor layout. The horizontal error (error of $x_k$) is proportional to the distance $r$, this is obviously not a desirable feature since under this layout, the ratio $\frac{r}{x}$ can be very large, so experiments need to be performed to verify the horizontal error, especially in cases of large $\frac{r}{x}$. Meanwhile, the vertical error (error of $y_k$) decreases when the ratio $\frac{r}{x}$ increases, when $r/x\to \infty$, the term $\frac{\partial \tilde{y_k}}{\partial e}\to 1$, the error will be the same as the ranging error $e$.

The above analysis shows that the horizontal error and the vertical error of the localization under the on-vehicle anchor layout can be very different. Both error terms are related to the ratio of the distance of the tag to the anchors and the distance between anchors, which is $r/x$.  The horizontal error is proportional to this ratio. Since in most applications, this ratio can be very large, the horizontal error can be hugely amplified from this layout. On the other hand, the vertical error is also related to this term, but since $\frac{x^2}{r^2}$ in most applications is very small, the vertical error should remain stable and be very similar to the ranging error $e$.

\subsection{Setup for experiments}
Tests were conducted to determine the coverage and evaluate accuracy of the distances measured by a system of three devices, each comprising a UWB module mounted on a LPCXpresso board - two serving as anchors placed at known locations with the third one as a tag to locate detectable points. Of the ranging modes offered by the device mentioned in an earlier section, DS ranging (Section \ref{ss:intro_dev}) were used for all the evaluations.  Throughout the tests, the anchors were fixed at pre-determined positions while the tag is moved to different locations. The experiments were carried out at an empty parking lot behind the Waterfront Shopping Mall in Pittsburgh, Pennsylvania, with no vehicular traffic or physical obstructions. 

\begin{figure} [ht]
    \centering
  \subfloat[Anchor mounted on the rear-view mirror\label{fi:anchor_mount}]{%
       \includegraphics[width=0.55\linewidth]{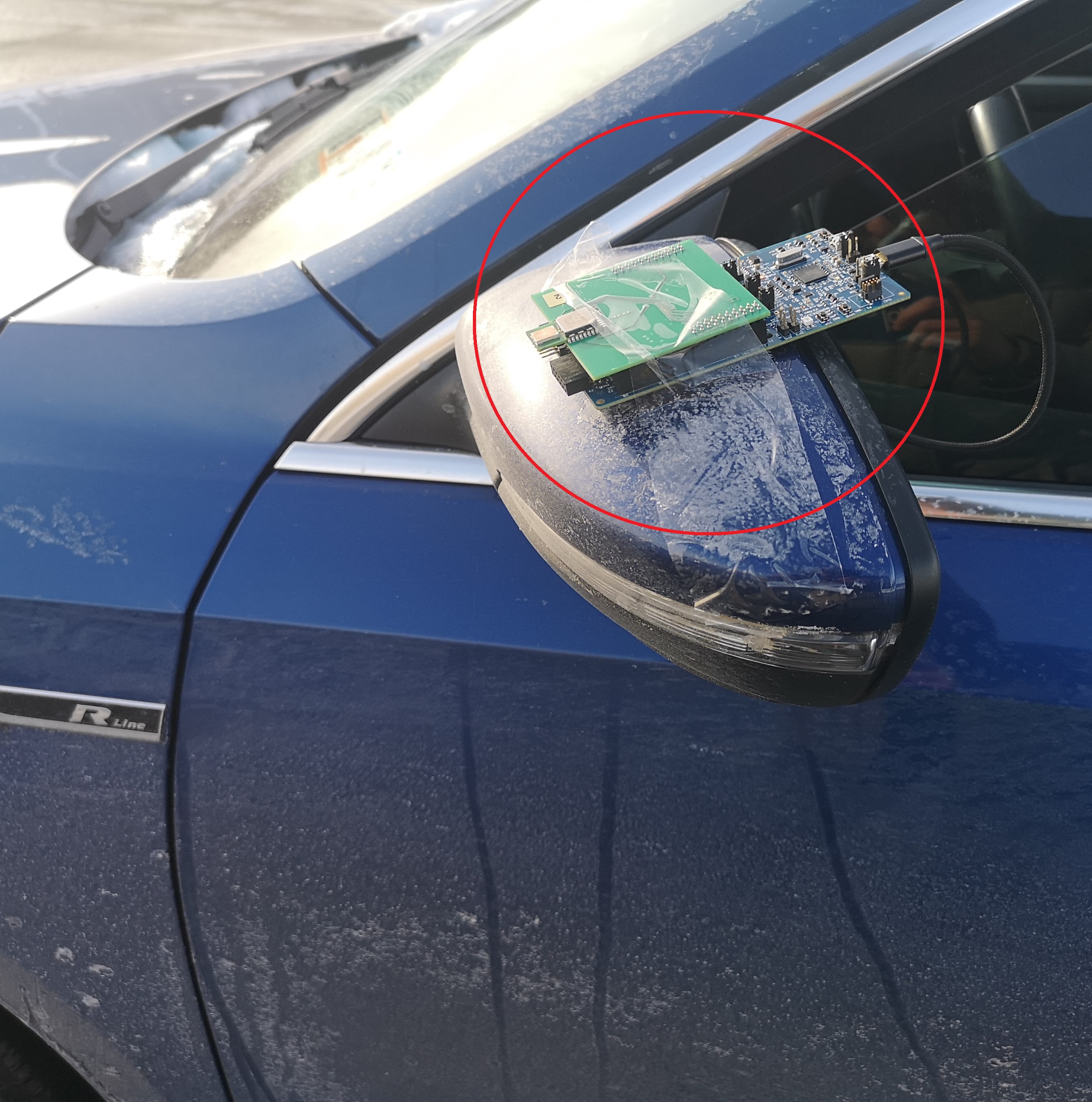}}
    \hfill
  \subfloat[Tag mounted on a tripod\label{fi:tag_mount}]{%
        \includegraphics[width=0.4\linewidth]{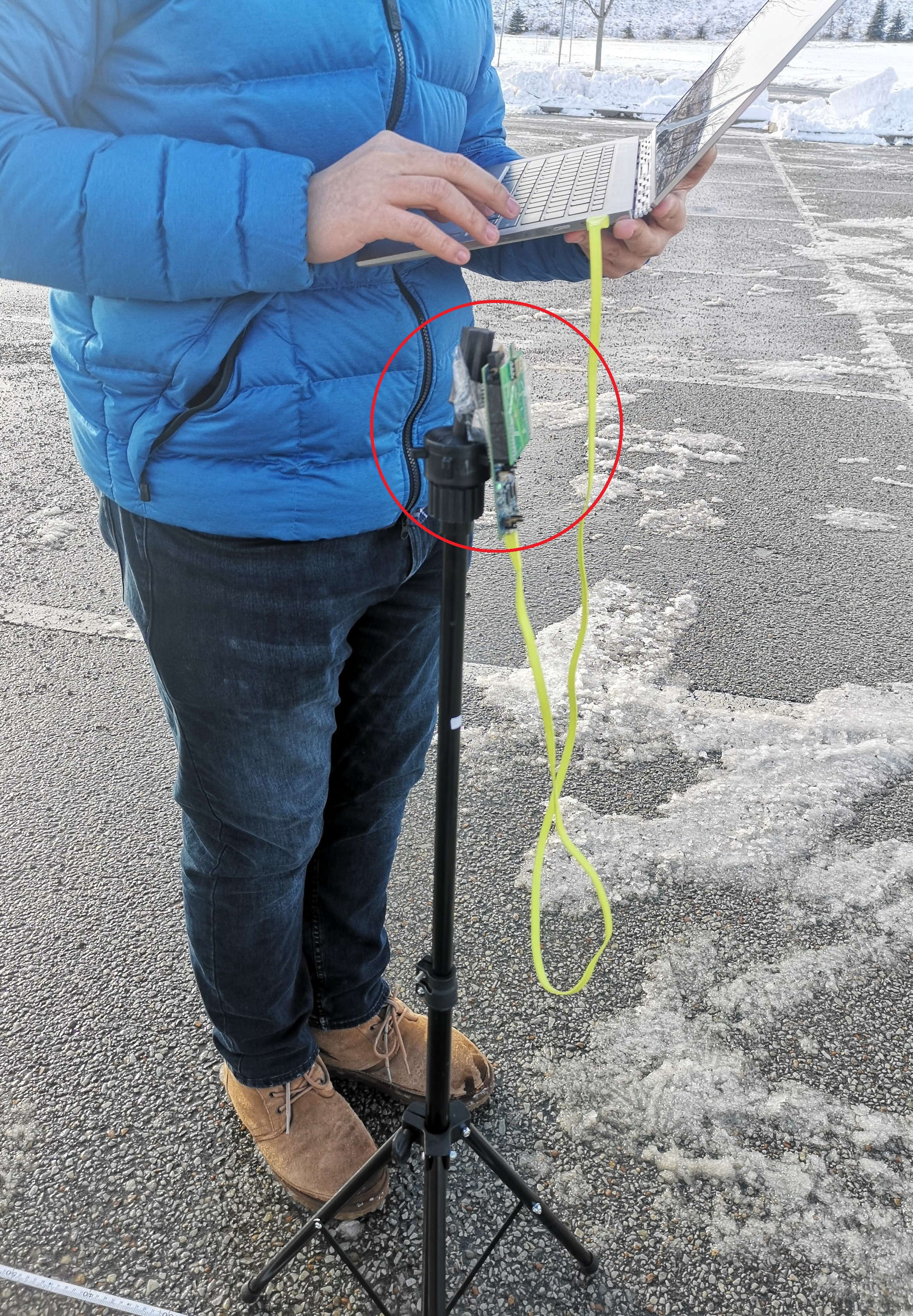}}

  \caption{Photos of anchor and tag used in the experiments}
  \label{fi:mount} 
\end{figure}

Figure \ref{fi:mount} shows the anchor and tag in the experiments. The anchors are mounted on the rear-view mirrors of the vehicle while the tag is mounted on a tripod.  The tag and the anchors are all mounted at a 1.1-meter height above the  ground. The distance between the two anchors is 1.85 meters ($x = 1.85/2$ merters in equation (\ref{eq:t1}) ). 

\subsection{Field Coverage}
For this particular experiment, field coverage implies the area  detected by both anchors. After mounting the anchors, the tag is moved to different locations to see if it can successfully obtain the distances to both of the anchors. To get realistic coverage of the on-vehicle anchor layout, we cover the tag with a thick puffer jacket, since tags are often carried in pockets or bags. The transmit power density is set to be -41.3 dBm/MHz (as suggested in the manual).

\begin{figure}[ht]
    \centering
    \includegraphics[width=.6\linewidth]{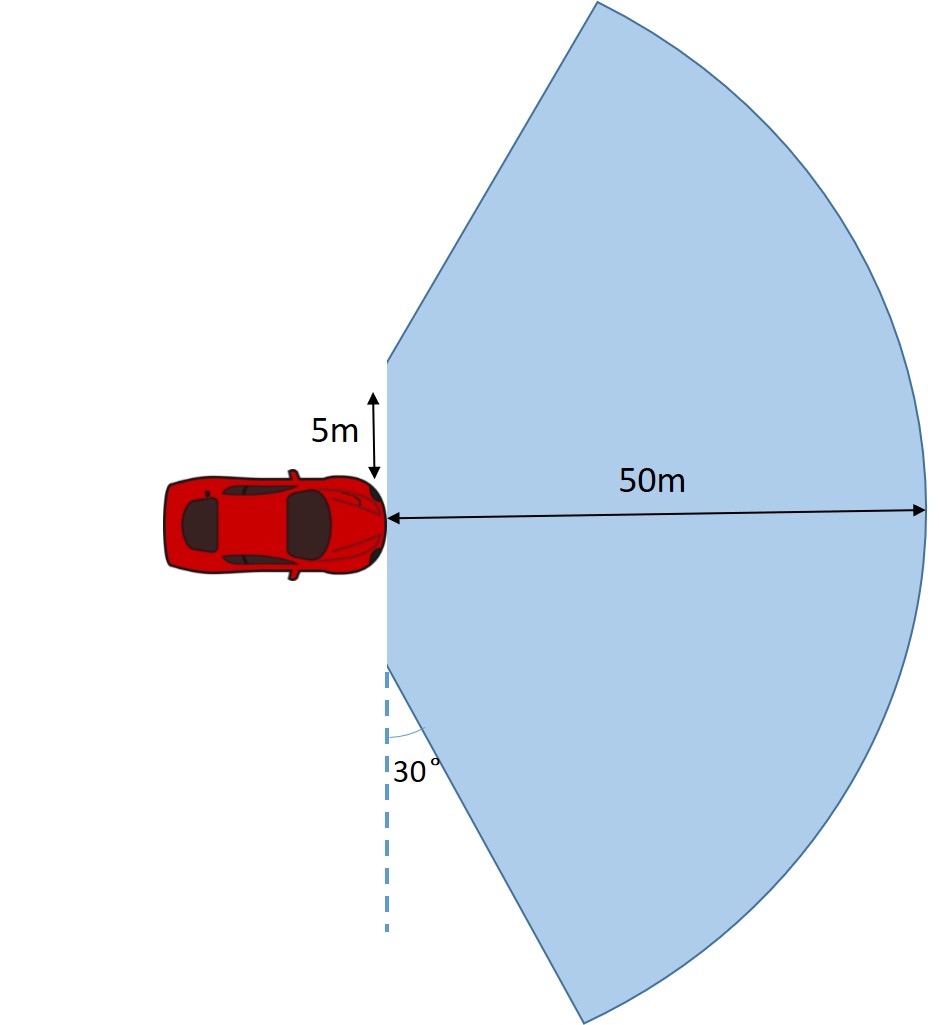}
    \caption{Coverage of the on-vehicle anchor layout, the portion of the figure is adjusted for better visualization (the vehicle should be much smaller and the sector range should be much bigger)}
    \label{fi:coverage}
\end{figure}

Figure \ref{fi:coverage} shows the coverage of the on-vehicle anchor layout in front of the vehicle. The two anchors can get the distance to the tag while the tag is more than 50 meters away from the vehicle in front. The viewing angle in front is roughly 120 degrees. One of the anchors loses signal from the tag when the tag is within the 30-degree region on the side (shown in  Figure \ref{fi:coverage}) because the vehicle blocks the line-of-sight, but the other anchor can still determine the distance from the tag. Though the coverage on the side of the vehicle significantly degrades, both anchors can detect the tag on the side within 5 meters.  The coverage of this layout is adequate for many applications aiming at locating the tags in front of the vehicle.


\subsection{Error characteristics}

In this subsection, we report the localization error performance. The experiment is performed in the same parking lot, there are very few random NLOS effect, so all the systematic errors under this setting can be calibrated, and we essentially account for the random error. Therefore, the localization error is evaluated by doing a batch of measurement and calculating the standard deviation of all points after triangulation.

\begin{figure}[ht]
    \centering
    \includegraphics[width=.9\linewidth]{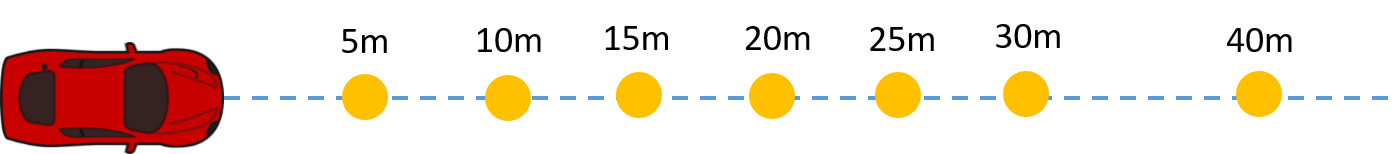}
    \caption{Locations chosen to test the error of localization (yellow dots), all points are in front of vehicle at distances of 5m, 10m, 15m, 20m, 25m, 30m and 40m, respectively.}
    \label{fi:exp2_setup}
\end{figure}

According to Section \ref{ss:error_analysis}, the horizontal error and vertical error behave very differently. Therefore, we measure the horizontal error and vertical error separately. The distance of the tag to the vehicle can be a significant factor affecting the horizontal error,  the experiment is repeated at different distances in front of the vehicle.

Figure \ref{fi:exp2_setup} shows the locations of error testing. Since front of the vehicle is of most interest, we chose locations in front of the vehicle with different distances up to 40 meters. At each location, 200 measurements were taken. We then do triangulation for each measured point to get Cartesian Coordinates $(x_i,y_i)$. The standard deviation of  $\{x_i\}$ and $\{y_i\}$ is calculated as an indicator of the horizontal and vertical error.

\begin{figure}[ht]
    \centering
    \includegraphics[width=.8\linewidth]{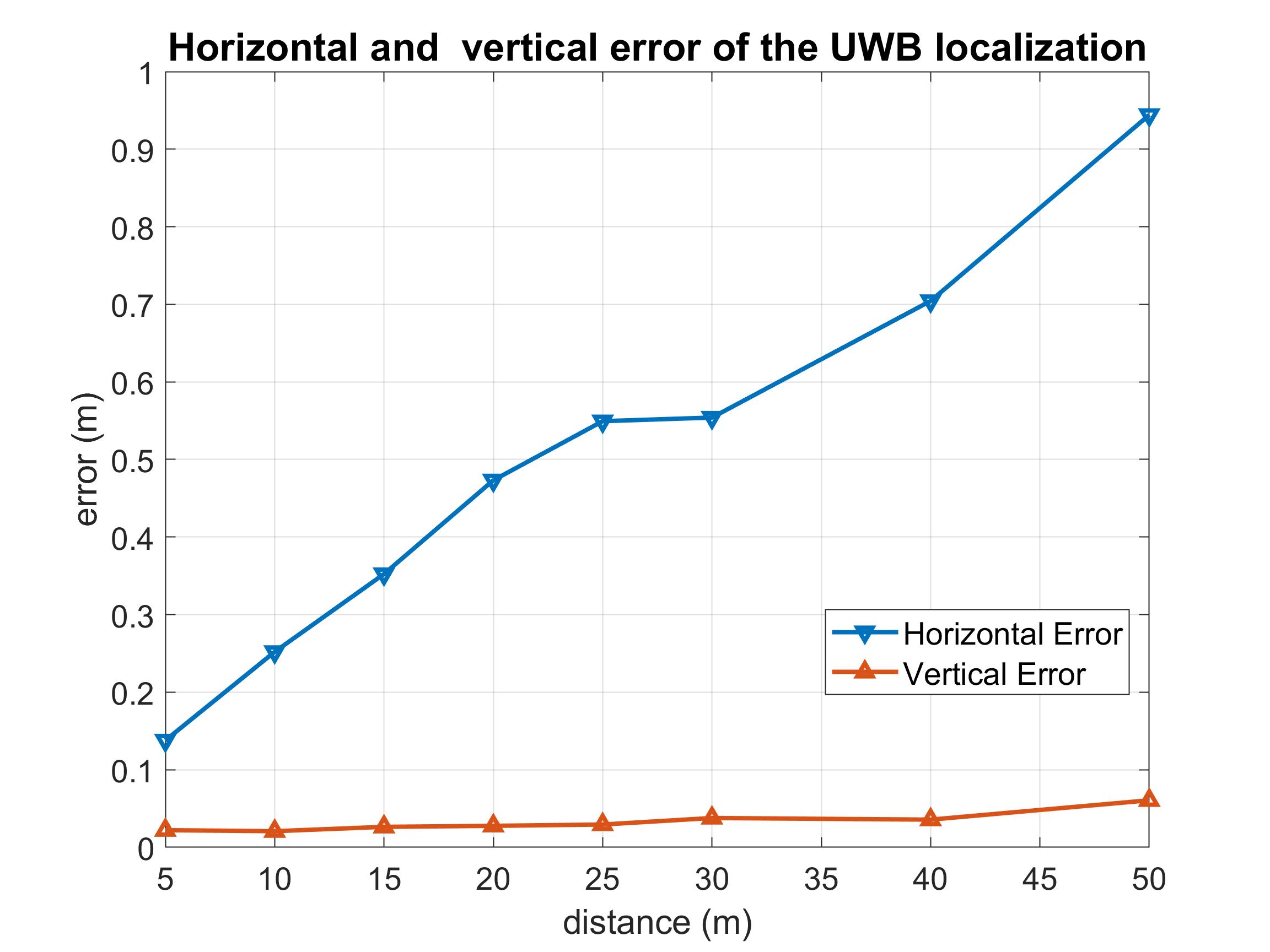}
    \caption{Localization error at different distances}
    \label{fi:exp2_result}
\end{figure}

Figure \ref{fi:exp2_result} shows the resulting measurement errors at different distances. The blue line shows the horizontal error, and the orange line shows the vertical error. We can note that, in general, the horizontal error shows quite good linearity. Meanwhile, the vertical error stays low and stable. Both trends are predicted accurately by the analysis outlined in section \ref{ss:error_analysis}. 

The horizontal error is much larger than the vertical error, and it reaches around 1 meter at a distance of 50 meters. The sizeable horizontal error may be the bottleneck of this layout. However, one can take an average of consecutive measurements, or apply a Kalman Filter to reduce the random error.

\section{Virtual Pedestrian Traffic Lights}
In this section, we introduce an application designed to utilize the on-vehicle anchor layout, known as Virtual Pedestrian Traffic Lights (VPTL). The application is designed under the futuristic assumption that future cellphones and wearable devices will be equipped with UWB chips.. 

The VPTL system is designed to be a subsystem of Virtual Traffic Lights (VTL) \cite{ferreira2010self, tonguzred, tonguz2019harness},  an infrastructure-free traffic lights system. The first VTL system prototype was built and publicly demonstrated in Riyadh, Saudi Arabia, in July 2018 \cite{zhang2018virtual, zhang2018increasing}.

\subsection{System Design}
The VTL algorithm has been discussed in \cite{ferreira2010self}. To briefly recap, the principle of operation is as follows:
\begin{enumerate}
\item \label{VTLStep_sensing} \emph{Sensing:} Vehicles approaching the intersection start to detect other vehicles via V2V communications.
\item \label{VTLStep_election} \emph{Leader Election:} If conflict is detected, vehicles select a leader which temporarily serves as a traffic light.

\item \label{VTLStep_broadcast} \emph{Broadcast:} The leader elected from the last step broadcasts traffic light information which gives red phase to its own lane and green phase to the lanes in the orthogonal direction.

\item \label{VTLStep_handover} \emph{Handover:} After the leader is elected, it decides how long each direction should receive the right-of-way. 

\item \label{VTLStep_release} \emph{Release:} When the leader doesn't detect any conflicting vehicle, it will give green to its own lane and then release the leadership functionality at the intersection. Now the intersection has no leader, and whenever there is a new conflict, vehicles will re-elect a leader, starting from step 1.

\end{enumerate}

To adopt the VPTL system, some minor modifications will be needed:
\begin{enumerate}
\item In the \emph{Leader Election} phase, the conflict will not only include vehicle-vehicle conflict, but also vehicle-pedestrian conflict.
\item In the \emph{Handover} step, if VPTL is needed, instead of handing over to another VTL leader, the current VTL leader will hand over to the VPTL leader and the VPTL leader will perform a pedestrian phase. 
\item The VPTL leader will track pedestrians at the intersection who are willing to cross the street, if all pedestrians have crossed the street, it will handover to the next VTL leader according to the VTL \emph{Handover} step; then, the VTL will return to its normal process. If not all of the pedestrians can be tracked, it will keep the pedestrian phase for a fixed time period.
\end{enumerate}
\begin{figure}[ht]
\centering
\includegraphics[width=3in]{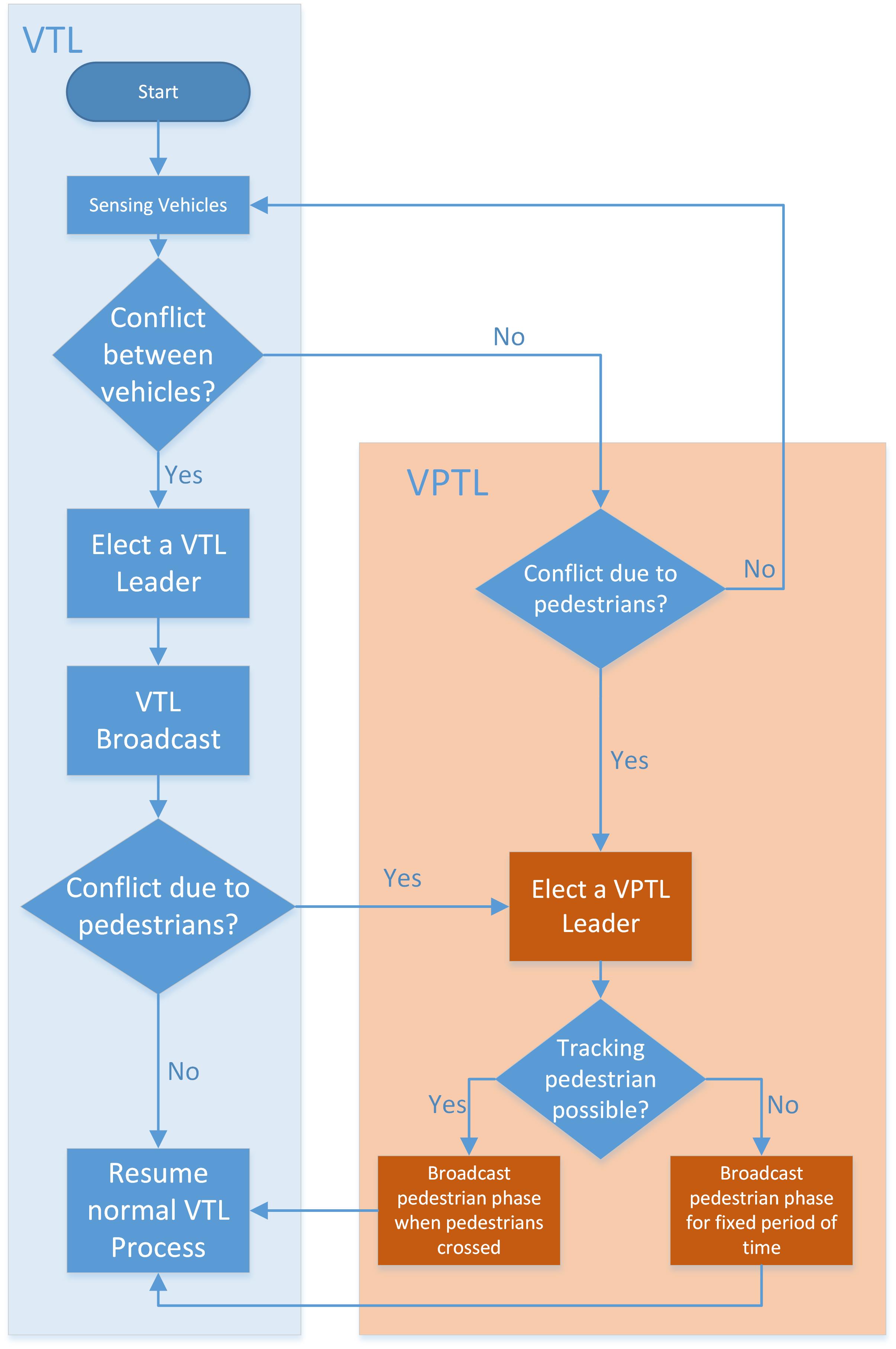}
\caption{Flow chart of VPTL system and how it interacts with the original VTL system}
\label{fi:algo}
\end{figure}

Figure \ref{fi:algo} shows how the VPTL system interacts with the original VTL system. While the original VTL logic is shown using a  blue color flowchart, the VPTL logic is shown using an orange color flowchart. In the figure, the blue components are the existing components in VTL logic, the orange ones are the new components needed to embed the VPTL subsystem into the original system. Notice that the VTL flowchart in blue color in Figure \ref{fi:algo}  is simplified and some details are omitted to avoid unnecessary complication.

\subsection{Field Tests}
We have performed a field test to evaluate the feasibility of the proposed system. The goal of this field test is to check if the on-vehicle anchor layout would be suitable for the localization requirements of VPTL system.  Namely, we want to verify that the on-vehicle anchor layout can distinguish if the pedestrian has crossed the street.

\begin{figure}[ht]
    \centering
    \includegraphics[width=.9\linewidth]{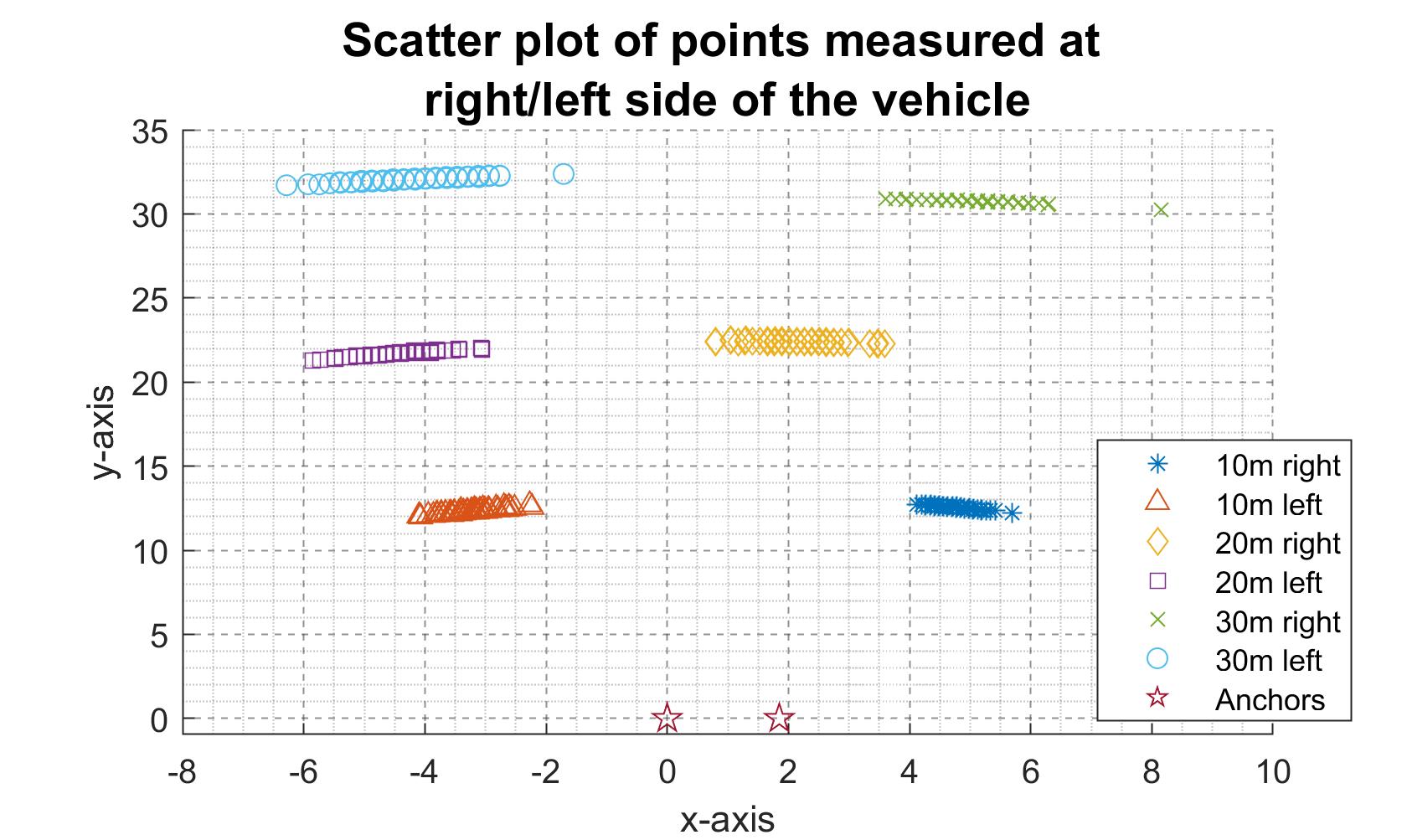}
    \caption{Scatter plot of points on left/right side of the vehicle, at the distance of 10m, 20m, and 30m}
    \label{fi:exp3_result}
\end{figure}

We take 200 measurements on both sides of the vehicle (approximately 5 meters to the side) at three different distances: 10m, 20m, and 30m. Figure \ref{fi:exp3_result} shows the scatter plot of all the points. From the figure, it is clear that the points on the left-hand side of the vehicle are entirely separated from the points on the right-hand side of the vehicle. Therefore, one can quickly check if the pedestrian has crossed from the left-hand side of the vehicle to the right-hand side of the vehicle, using methods such as hypothesis testing or support vector machine. Our future research will investigate these different approaches.

The raw scatter image clearly shows that the designed system would work  for the VPTL application. For some other applications requiring better accuracy, a Kalman Filter can be applied to reduce the error further.  Implementation of Kalman Filter is ongoing work that will be reported elsewhere.

\section{Conclusions}
In this paper, we examine the feasibility of using UWB technology in vehicles for localization with an anchor layout. The benefit of our approach is that it allows  vehicles to locate a tag in an infrastructure-free manner.  Theoretical and experimental results are reported to quantify the error characteristics of such an anchor installation layout. The experiments show that the horizontal error can be as large as 1 meter, which is adequate for some applications. However, the sizeable horizontal error might be the bottleneck for some accuracy-sensitive applications. Additional research is required to reduce the error measured in our field tests.

An interesting illustrative application for the presented layout using UWB technology, Virtual Pedestrian Traffic Light (VPTL), is also reported in the latter part of the paper. Experiments show that the accuracy of the on-vehicle anchor layout is pretty good and could be adequate for the VPTL application.  This interesting application paves the way for enhancing the VTL algorithm to include pedestrian  detection and localization at intersections.







\bibliographystyle{IEEEtran}
\vspace{-0.05in}
\bibliography{IEEEabrv,reference}
%



\end{document}